\documentclass[12pt]{iopart}
\usepackage{graphicx}
\usepackage{iopams}

\begin{document}

\title{Gating of high-mobility two-dimensional electron gases in GaAs/AlGaAs
heterostructures.}

\author{C. R{\"{o}}ssler}
\address{Solid State Physics Laboratory, ETH Zurich, 8093
Zurich, Switzerland}
\ead{roessler@phys.ethz.ch}

\author{T. Feil}
\address{Solid State Physics
Laboratory, ETH Zurich, 8093 Zurich, Switzerland}

\author{P. Mensch}
\address{Solid State Physics Laboratory, ETH Zurich, 8093
Zurich, Switzerland}

\author{T. Ihn}
\address{Solid State Physics Laboratory, ETH Zurich, 8093
Zurich, Switzerland}

\author{K. Ensslin}
\address{Solid State Physics Laboratory, ETH Zurich, 8093
Zurich, Switzerland}

\author{D. Schuh}
\address{Institut f{\"{u}}r Experimentelle und Angewandte Physik, Universit{\"{a}}t Regensburg, 93040 Regensburg, Germany}

\author{W. Wegscheider}
\address{Solid State Physics Laboratory, ETH Zurich, 8093
Zurich, Switzerland}

\begin{abstract}
We investigate high-mobility two-dimensional electron gases in
$\rm{Al}_x \rm{Ga}_{1-x} \rm{As}$ heterostructures by employing
Schottky-gate-dependent measurements of the samples' electron
density and mobility. Surprisingly, we find that two different
sample configurations can be set in situ with mobilities differing
by a factor of more than two in a wide range of densities. This
observation is discussed in view of charge redistributions between
the doping layers and is relevant for the design of future gateable
high-mobility electron gases.
\end{abstract}

\maketitle

\section{Introduction}
Two-dimensional electron gases (2DEGs) in $\rm{Al}_x \rm{Ga}_{1-x}
\rm{As}$ heterostructures can reach mobilities exceeding $10^7\,
\rm{cm^2/Vs}$ at low temperatures, facilitating the observation of
fascinating phenomena like the microwave-induced zero-resistance
states, the $\nu=5/2$ quantum Hall state, and interactions between
composite fermions~\cite{pfe03}. It turns out that to this end,
clean materials and growth techniques beyond the standard modulation
doping approach are essential~\cite{fri96,uma97,hwa08}. However,
current growth protocols are apparently in conflict with the in situ
control of the 2DEG's density via Schottky-gates: recent experiments
on gated high-mobility structures have reported hysteresis effects
and temporal drifts when biasing Schottky gates \cite{dol08,mil07}.
In order to investigate the origin of these effects, we fabricate
gated Hall bars containing a high-mobility 2DEG.
\section{Experiment}
In the investigated samples, the 2DEG resides in a $30\,\rm nm$ wide
GaAs quantum well, $160\,\rm{nm}$ beneath the surface. The overall
growth scheme is similar as previously reported \cite{uma09}. Two
Si-donor layers are situated at depths $z$ of about $40\,\rm{nm}$
and $680\,\rm{nm}$ in order to compensate for surface-states and
background impurities in the substrate. Since these donors are
embedded in $\rm{Al}_{0.33} \rm{Ga}_{0.67} \rm{As}$, they are
expected to form DX-centers below the $\Gamma$-band~\cite{moo90}.
Layers of GaAs at $z\approx70\,\rm{nm}$ and $z\approx250\,\rm{nm}$
are $\delta$-Si doped and enclosed by $2\,\rm{nm}$ thick layers of
AlAs. The ground state in the AlAs wells ($X$-minima) is lower than
that in the GaAs wells ($\Gamma$ minima). Consequently, only a part
of the donors' free electrons reside in the high-mobility 2DEG,
whereas the rest is expected to populate the $X$-band within the
AlAs-layers embedding the doping planes~\cite{fri96,uma09}.

The investigated samples stem from different wafers with the same
layer structure. Typical transport characteristics of unprocessed
samples are a Drude mobility $\mu\gtrsim18\times10^6\,\rm cm^2/Vs$
at an electron density $n_{\rm S}=3.2\times10^{11}\,\rm cm^{-2}$ as
measured in van-der-Pauw geometry at a temperature of $T=1.3\,\rm K$
(data not shown).

The samples are etched $150\,\rm{nm}$ deep to define a
$1100\,\rm{\mu m}\times100\,\rm{\mu m}$ wide Hall bar. AuGe pads are
deposited and annealed to contact the 2DEG. In a final step,
$200\,\rm{nm}$ of Ti/Au are deposited on the Hall bar to form a
Schottky top-gate.

\begin{figure}
\includegraphics[scale=1]{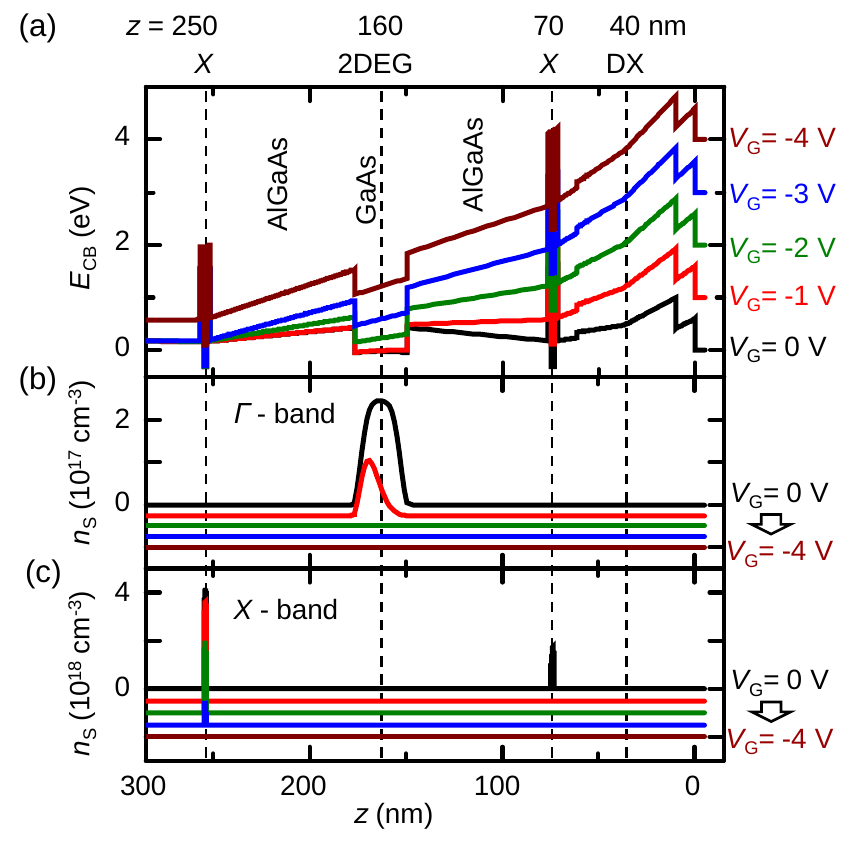}
\caption{\label{fig:bandstructure} (colour online) (a) Calculated
conduction band profile of the investigated heterostructure, shown
for top-gate voltages $V_{\rm G}=0,-1,-2,-3,-4\,\rm V$ (black to
maroon). The gate is situated at $z\leq0\,\rm{nm}$, Si-dopants in
AlGaAs at $z\approx40\,\rm{nm}$ and $z\approx680\,\rm nm$ compensate
for surface states and substrate impurities, respectively.
Si-dopants in GaAs at $z\approx70\,\rm nm$ and $z\approx250\,\rm nm$
are bordered by AlAs layers with the well state in the $X$-band (not
shown) situated below the Fermi energy (defined as $E_{\rm
CB}=0\,\rm eV$) for $V_{\rm G}=0\,\rm V$. The 2DEG resides in a GaAs
quantum well at $z\approx160\,\rm{nm}$. (b) Calculated free electron
density in the $\Gamma$-band for $V_{\rm G}=0,-1,-2,-3,-4\,\rm V$,
with curves vertically offset for clarity. With increasingly
negative gate bias, the 2DEG's density decreases until it is fully
depleted at $V_{\rm G}=-2\,\rm V$ (green trace). (c) Electron
density in the $X$-band, calculated for the same set of gate
voltages. The upper layer of $X$-electrons is depleted first, then
the lower layer of $X$-electrons is depleted at $V_{\rm G}\lesssim
-3\,\rm V$.}
\end{figure}

Figure~\ref{fig:bandstructure} (a) shows the uppermost $300\,\rm nm$
of the heterostructure's $\Gamma$ conduction band edge, calculated
self-consistently for different top gate voltages~\footnote{We
employ the 3D nanodevice simulator $\rm{nextnano}^3$, available at
http://www.nextnano.de}. The electrochemical potential (defined as
$E_{\rm CB}=0\,\rm eV$) is assumed to be pinned mid-gap in the
substrate and $0.6 \,\rm eV$ below the $\Gamma$ conduction band at
the surface. For the sake of simplicity, dynamic processes are not
taken into account and the ionisation degree of the DX-dopants
dopants is assumed to be independent of gate voltage. The doping
concentrations at $z = 70\,\rm nm$ and $z = 250\,\rm nm$ have been
chosen so that both the 2DEG's electron density at $V_{\rm G} =
0\,\rm V$ and the pinch-off voltage are consistent with the
experimentally determined values. The chosen value of
$8.0\times10^{11}\,\rm cm^{-2}$ is smaller than the density of
silicon atoms of $20\times10^{11}\,\rm cm^{-2}$, suggesting that
only part of the donors are electrically active.
The gate is situated at the surface ($z=0\,\rm nm$) and is set to
$V_{\rm G}=0,-1,-2,-3,-4\,\rm V$, respectively (black, red, green,
blue and maroon trace). Figure~\ref{fig:bandstructure} (b) shows the
calculated free electron density in the $\Gamma$-band for these gate
voltages. Applying $V_{\rm G}=-1\,\rm V$ reduces the 2DEG's electron
density and shifts the wave function to the lower boundary of the
well (red trace). The 2DEG is depleted when applying $V_{\rm
G}=-2\,\rm V$ (green trace). Figure~\ref{fig:bandstructure} (c)
shows the electron density in the $X$-band for the same set of gate
voltages. At $V_{\rm G}=0\,\rm V$ (black trace), electrons occupy
the AlAs layers at $z\approx70\,\rm{nm}$ and $z\approx250\,\rm{nm}$
with the respective electron densities in the $X$-band being $n_{\rm
S}=2.1\times10^{-11}\,\rm cm^{2}$ (upper layer) and $n_{\rm
S}=5.5\times10^{-11}\,\rm cm^{2}$ (lower layer). Applying $V_{\rm
G}=-1\,\rm V$ (red) depletes the upper layer of $X$-electrons and
$V_{\rm G}\lesssim-3\,\rm V$ depletes also the bottom layer of
$X$-electrons.

The samples' transport properties are measured at low temperatures
$T\lesssim4.2\,\rm{K}$ employing a standard lock-in technique with a
typical modulation frequency of $f=72\,\rm{Hz}$ and an amplitude
$\Delta V_{\rm SD}=40\,\rm{\mu V}$. The Hall density $n_{\rm S}$ is
obtained from the measured source-drain current $I_{\rm SD}$ and the
transversal four terminal voltage $U_{Y}$ via the relationship
$n_{\rm S}=BI_{\rm SD}/{\rm e}U_{Y}$. Here, $B$ is the magnetic
field strength applied perpendicular to the plane of the 2DEG and
${\rm e}$ denotes the elementary charge. Starting from positive gate
bias $V_{\rm G}=+0.5\,\rm{V}$, the gate voltage is swept with a rate
of $\Delta V_{\rm G}/\Delta t\approx 10\,\rm mV/s$ to $V_{\rm
G}=-3\,\rm{V}$ and back to $V_{\rm G}=+0.5\,\rm{V}$. The upper limit
of $V_{\rm G}=+0.5\,\rm{V}$ is determined by the onset of leakage
current, whereas the lower boundary of $V_{\rm G}=-3\,\rm{V}$ was
empirically chosen because larger voltage loops do not change the
results described in the following. The calculated electron
densities shown in figure~\ref{fig:bandstructure} (c) suggest that
this lower border is correlated with the depletion of the lower
layer of $X$-electrons for $V_{\rm G}\lesssim-3\,\rm V$.

\begin{figure}
\includegraphics[scale=1]{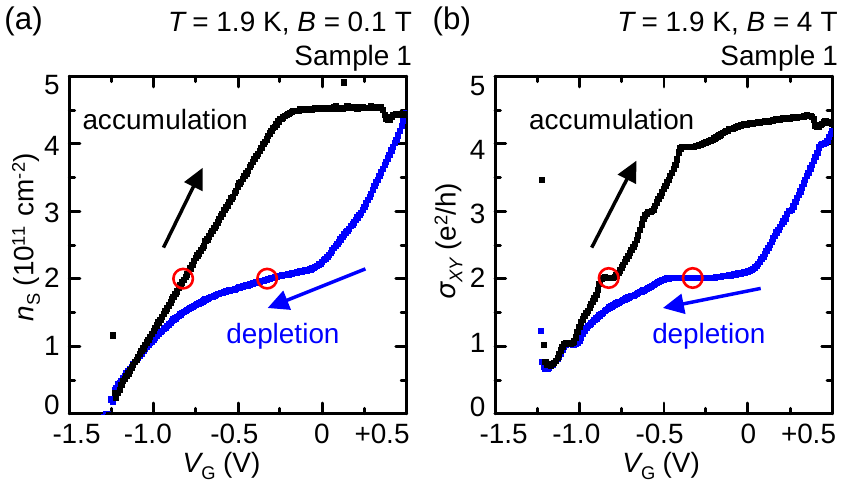}
\caption{\label{fig:density} (colour online) (a) Electron density
$n_{\rm S}$ as a function of gate voltage, measured at small
perpendicular magnetic field $B=0.1\,\rm T$ at $T=1.9\,\rm K$.
Sweeping from $V_{\rm G}=+0.5\,\rm{V}$ to $V_{\rm G}=-3\,\rm{V}$
(blue), the density is reduced in a nonlinear manner. In contrast,
the density increases linearly while going to more positive gate
bias (black). The gate voltages corresponding to a density of
$n_{\rm S}=2\times10^{11}\,\rm cm^{-2}$ are marked by red circles.
(b) Transversal magnetoconductance $\sigma_{XY}=I_{\rm SD}/U_{Y}$ in
units of $\rm e^2/h$, plotted as a function of the applied gate bias
at $B=4\,\rm T$. The two gate voltages with $\sigma_{XY}=2\,\rm
e^2/h$ and hence filling factor $\nu=2$ are again marked by red
circles.}
\end{figure}

Figure~\ref{fig:density} (a) shows the measured electron density as
a function of the gate bias in magnetic field $B=0.1\,\rm T$
oriented perpendicular to the plane of the 2DEG.
Clearly, there is a strong hysteresis between depletion (blue trace)
and accumulation of the 2DEG (black). For comparison, two red
circles highlight the gate voltages where $n_{\rm
S}=2\times10^{11}\,\rm cm^{-2}$. Figure~\ref{fig:density} (b) shows
a similar measurement repeated at $B=4\,\rm T$, where the density is
expressed as the transversal magnetoconductance $\sigma_{XY}=I_{\rm
SD}/U_{\rm Y}$ in units of $\rm e^2/h$. Having entered the quantum
Hall regime, plateaus of $\sigma_{XY}$ correspond to integer filling
factors $\nu=n_{\rm S}{\rm h/e}B$ and hence $\nu=n_{\rm
S}\times1.03\times10^{-11}\,\rm cm^{2}$. Again, gate voltages
corresponding to the electron density $n_{\rm S}=2\times10^{11}\,\rm
cm^{-2}$ are marked by red circles. These voltages match the values
from Figure~\ref{fig:density} (a), implying that the classical Hall
density matches the quantum Hall density and no populated second
subband or other parallel conducting channels contribute to the
measured electron densities within experimental accuracy, as
confirmed by zero resistance minima in Shubnikov-de-Haas data (not
shown). For both measurements, as long as the 2DEG density is being
reduced, $n_{\rm S}$ is not linear with respect to the gate bias. In
contrast, the electron density increases linearly with a slope of
$dn_{\rm S}/dV_{\rm G}=4.2\times10^{11}\,\rm{cm^{-2}V^{-1}}$ while
accumulating the 2DEG, being in good agreement with the model of a
parallel-plate capacitor defined by the gate and the
2DEG~\footnote{Assuming the plates to be formed by the top gate and
the 2DEG at a depth of $d\approx160\,\rm{nm}$, the expected slope
$dn_{\rm S}/dV_{\rm G}$ is given by $dn_{\rm S}/dV_{\rm
G}=C/\rm{e}=\rm{\varepsilon\varepsilon}_0/\rm{e}\textit{d}\approx4.2\times10^{11}\,\rm{cm^{-2}V^{-1}}$,
where $C$ is the capacity between top gate and 2DEG,
$\varepsilon\approx12$ the dielectric constant of $\rm{Al}_{0.33}
\rm{Ga}_{0.67} \rm{As}$ and
$\varepsilon_0=8.85\times10^{-12}\,\rm{CV^{-1}m^{-1}}$ is the
electric field constant.}. This qualitative difference indicates
that charge redistributes in the region between gate and 2DEG during
depletion but not during accumulation of the 2DEG.

\begin{figure}
\includegraphics[scale=1]{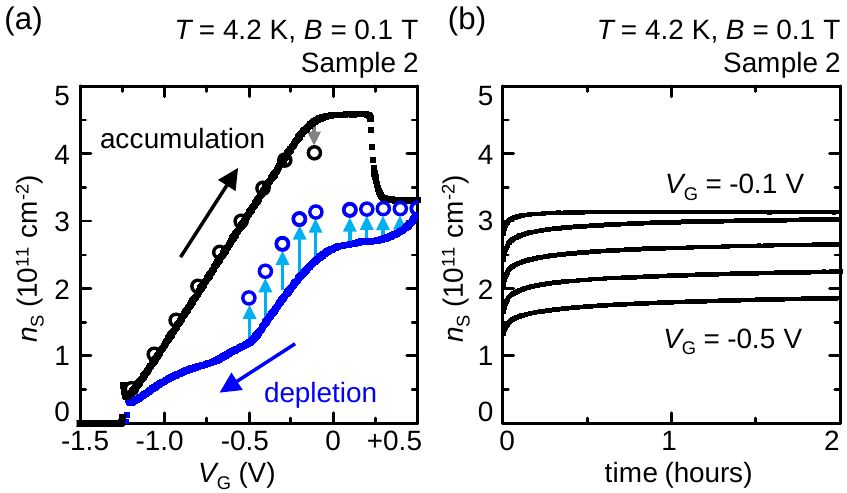}
\caption{\label{fig:time} (colour online) (a) Electron density
$n_{\rm S}$ as a function of gate voltage, measured during depletion
(blue) and during accumulation (black) of the 2DEG. When stopping
the gate-sweep during depletion, the electron density converges
towards the values shown as blue circles. A stop during accumulation
and waiting for one hour results in the densities shown as black
circles. During depletion, the sample's electron density drifts
towards larger values, whereas during accumulation the sample it is
stable for densities $n_{\rm S}\lesssim3.5\times10^{11}\,\rm
cm^{-2}$. (b) Time dependence of the electron density after the
depleting has been interrupted at one of the gate voltages $V_{\rm
G}=-0.1,-0.2,-0.3,-0.4,-0.5\,\rm{V}$.}
\end{figure}

Figure~\ref{fig:time} (a) shows the electron density of another
sample with the same heterostructure layer sequence, measured at
$B=0.1\,\rm T$ and temperature $T=4.2\,\rm K$.
Again, the electron density shows a strong hysteresis between
depletion (blue) and accumulation (black) of the 2DEG. Interrupting
the gate-sweep during depletion and waiting for two hours results in
an increased electron density (blue circles). In contrast, the same
procedure during accumulation shows that the electron density is
stable for at least one hour during accumulation (black circles).
The finding implies that the charge redistribution is complete at
gate biases $V_{\rm G}\lesssim-1.5\,\rm V$ and is stable in time
afterwards. As a comparison, the temporal evolution of the electron
density during depletion is shown in figure~\ref{fig:time} (b).
Stopping the gate-sweep at one of the gate voltages $V_{\rm
G}=-0.1,-0.2,-0.3,-0.4,-0.5\,\rm{V}$ results in a slow recovery of
the electron density.

A probable origin of the observed charge redistributions is the
charging of the $\delta-\rm doping$ layer $z\approx40\,\rm{nm}$
beneath the surface and/or the $X$-electrons at
$z\approx70\,\rm{nm}$. Thus, the observed hysteresis resembles the
memory-like behaviour in 2DEG samples with vertically tunnel coupled
self-assembled quantum dots~\cite{kan07,mar09}. The layers would
build up positive charge during depletion up to full ionisation at
very negative gate voltages $V_{\rm G}\lesssim-1.5\,\rm V$.
According to the calculation of the band structure in
figure~\ref{fig:bandstructure} (a), electrons would not be able to
repopulate the layers unless a sufficiently positive gate bias is
applied. In the gate-sweeps in figures~\ref{fig:density} and
\ref{fig:time}, such a reset of the 2DEG's density is indeed
observed when ramping the gate voltage up to $V_{\rm G}\sim0\,\rm
V$. The more drastic reset at $T=4.2\,\rm K$ (figure~\ref{fig:time})
indicates that the time constant of the associated charge
redistribution is temperature dependent.

\begin{figure}
\includegraphics[scale=1]{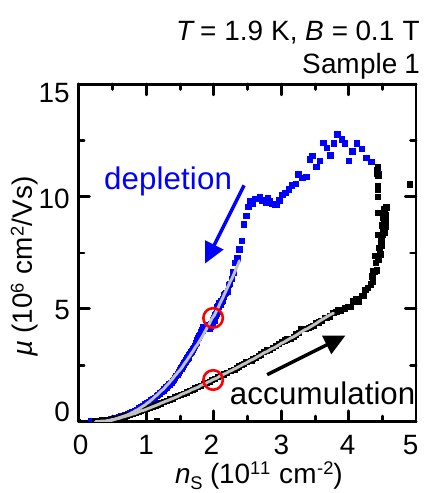}
\caption{\label{fig:mobility} (colour online) Drude mobility $\mu$
as a function of the 2DEG's electron density $n_{\rm S}$ measured at
$B=0.1\,\rm T$, $T=1.9\,\rm K$. The mobility while depleting (blue)
is much higher than when accumulating (black). For example at
$n_{\rm S}=2\times10^{11}\,\rm cm^{-2}$ (red circles), the
mobilities are found to be $\mu=4.5\times10^{6}\,\rm cm^{2}/Vs$
(depleting) and $\mu=1.9\times10^{6}\,\rm cm^{2}/Vs$ (accumulating),
respectively. Grey lines are fits of a power-law dependence
$\mu=c\times(n_{\rm S}-n_{0})^k$ to the data yielding exponents of
$k=2.1$ (depleting) and $k=1.3$ (accumulating).}
\end{figure}

Different charge configurations of the doping layers are expected to
have a strong impact on the Drude mobility $\mu=I_{\rm SD}/({\rm
e}n_{\rm S}U_{X})\times l/w$ of the 2DEG, where $U_{X}$ is the
longitudinal potential drop and $l/w=6$ is the aspect ratio of the
Hall bar. Figure~\ref{fig:mobility} shows the mobility, plotted as a
function of the electron density $n_{\rm S}$ at $B=0.1\,\rm T$,
$T=1.9\,\rm K$ during a gate-sweep from $V_{\rm G}=+0.5\,\rm{V}$ to
$V_{\rm G}=-3\,\rm{V}$ and back to $V_{\rm G}=+0.5\,\rm{V}$.
While depleting, the mobility is clearly larger than during
accumulation at any given density (e.g.,
$\mu\approx4.5\times10^{6}\,\rm cm^{2}/Vs$ as compared to
$\mu\approx1.9\times10^{6}\,\rm cm^{2}/Vs$ at $n_{\rm
S}\approx2\times10^{11}\,\rm cm^{-2}$, red circles). When
accumulating, a recovery of the mobility sets in at electron
densities $n_{\rm S}\gtrsim4\times10^{11}\,\rm cm^{-2}$,
corresponding to gate voltages $V_{\rm G}\gtrsim0\,\rm V$.

For temperatures of $T\lesssim2\,\rm K$, we find that the mobility
is independent of temperature (measurement not shown), indicating
that phonon scattering is not relevant for the interpretation of
this gate-sweep. The fit of a power-law dependence
$\mu=c\times(n_{\rm S}-n_{\rm 0})^k$ to the data yields exponents of
$k=2.1$ (depleting) and $k=1.3$ (accumulating)~\footnote{The other
fitting parameters are $c=3.4\times10^{-18}\,\rm
cm^2V^{-1}s^{-1}(cm^2)^k$, $n_0=0.2\times10^{11}\,\rm cm^{-2}$
during depletion and $c=8.4\times10^{-9}\,\rm
cm^2V^{-1}s^{-1}(cm^2)^k$, $n_0=0.4\times10^{11}\,\rm cm^{-2}$
during accumulation}. Such dependence is commonly associated with
remote impurity scattering of ionised donor atoms~\cite{hir86} and
is in contrast to $k\sim0.7$ found in background-impurity-limited
2DEGs~\cite{uma97}.

From the strong positive $n_{\rm S}$-dependence of our sample's
mobility it can be concluded that the dominant scatterers are
characterized by a potential $V(r)$ whose Fourier transform $V(q)$
diminishes strongly when $q$ is increased. The significantly larger
exponent $k$ and the higher mobility during depletion suggest that
during depletion, the disorder potential is steeper in $q$-space,
corresponding to a smoother (long-range-)potential in real space.

\begin{figure}
\includegraphics[scale=1]{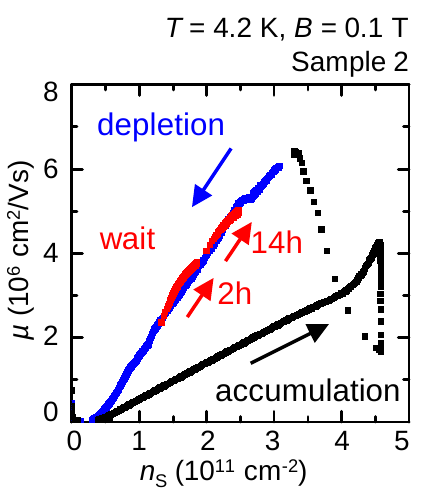}
\caption{\label{fig:mobility_wait} (colour online) Mobility $\mu$ as
a function of the 2DEG's electron density $n_{\rm S}$ measured at
$B=0.1\,\rm T$ and $T=4.2\,\rm K$. The mobility during depletion
(blue) is much higher than during accumulation (black). Stopping the
gate-sweep at $n_{\rm S}\approx1.3\times10^{11}\,\rm cm^{-2}$ and
waiting for two hours (lower red trace) results in a gradual
increase of the mobility and density along the depletion-trace. The
same behaviour is observed when stopping at $n_{\rm
S}\approx2\times10^{11}\,\rm cm^{-2}$ and waiting for 14 hours
(upper red trace). If the gate-sweep is stopped during accumulation,
the sample's density and mobility do not change over time.}
\end{figure}

The observed decrease of the mobility with decreasing electron
density could be caused by several effects:
\begin{enumerate}
\item Decreasing the electron density increases the Fermi-wavelength.
This in turn facilitates scattering events that arise from the
long-range part of the disorder potential~\cite{and82}.
\item $X$-electrons, which screen the dopants' disorder
potential at $V_{\rm G}=0\,\rm V$, are depleted at sufficiently
negative gate bias. The loss of the screening layer might have a
strong impact on the 2DEG's mobility.
\item During depletion, dopants between gate and 2DEG become ionised which leads to increased remote ionised impurity scattering.
\item At more negative gate biases (and hence lower densities), the confinement
potential of the 2DEG is more strongly tilted. In this situation,
the weight of the 2DEG's wave function is situated closer to the
lower AlGaAs-barrier, increasing alloy scattering and interface
roughness scattering~\cite{boc90}.
\end{enumerate}
The observation of different mobilities at the same density can not
be related to the first effect, because the Fermi wavelength is the
same in both cases. In order to test the impact of ionised donors
above the 2DEG onto the mobility, the temporal evolution of the
mobility is shown in Figure~\ref{fig:mobility_wait}.
The mobility is measured at a temperature of $T=4.2\,\rm K$
(compared to $T=1.9\,\rm K$ in Figure~\ref{fig:mobility}) in order
to speed up the relatively slow process of ionisation. As a
consequence, the dependence of the mobility on the density is almost
linear due to enhanced phonon scattering~\cite{hir86}, but still the
mobility is higher during depletion than during accumulation. During
depletion (blue), the gate-sweep is stopped and the change of the
sample's density and mobility are recorded. Starting from $n_{\rm
S}\approx1.3\times10^{11}\,\rm cm^{-2}$ (lower red trace) or $n_{\rm
S}\approx2.0\times10^{11}\,\rm cm^{-2}$ (upper red trace), both the
density and mobility increase over time. The evolution follows the
depletion curve, indicating that the progressing ionisation acts
primarily as a positive gate bias and does not determine the
mobility at a given electron density. This leads us to believe that
the drastic change of the mobility after fully depleting the 2DEG is
not due to changes of the charge states of the screening layer at
$z\approx70\,\rm nm$ or the doping layer at $40\,\rm nm$, but rather
due to the screening layer at $250\,\rm nm$. After depleting both
the 2DEG and this lower screening layer, more positive charge is
present beneath the 2DEG, increasing interface roughness scattering
and alloy scattering due to a more tilted confinement potential.
Additionally, the loss of this screening layer should increase the
2DEG's remote ionised impurity scattering with dopants beneath the
2DEG. The latter effect might be weaker than the effect of a tilted
potential, because the mobility shown in figure~\ref{fig:mobility}
does not show a drastic decrease at the gate voltage corresponding
to the expected depletion of the upper screening layer.

\begin{figure}
\includegraphics[scale=1]{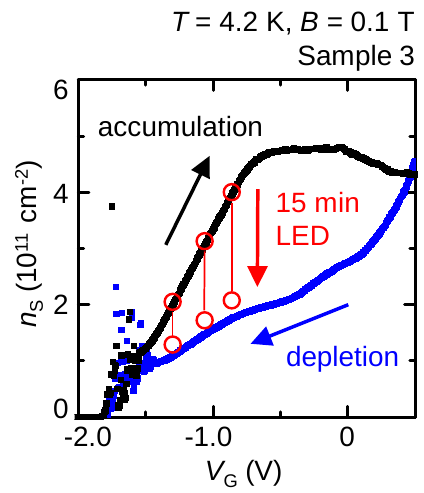}
\caption{\label{fig:led} (colour online) Electron density $n_{\rm
S}$ as a function of gate voltage, measured at $B=0.1\,\rm T$,
$T=4.2\,\rm K$. There is a strong hysteresis between depleting (blue
curve) and accumulating of the 2DEG (black curve). The vertical
(red) lines show the density change during $15\,\rm minutes$ of
IR-LED illumination with a current of $I=0.1\,\rm mA$ at gate
voltages of $V_{\rm G}=-1.3,-1.06,-0.86\,\rm V$. During
illumination, the density converges towards the value observed
during depletion.}
\end{figure}

The finding implies that the bottom screening layer cannot be
refilled laterally (from non-depleted areas of the Hall bar) but
requires vertical tunnelling or an additional activation energy to
regain its' original $X$-electron density. In order to test this
interpretation, the response of the electron density to illumination
with an infrared light emitting diode (IR-LED) is investigated.
According to the bandstructure calculation shown in
figure~\ref{fig:bandstructure} (a), the conduction band edge of the
doping layers between top gate and 2DEG is above the Fermi energy
for gate voltages $V_{\rm G}\lesssim-0.5\,\rm V$. Hence, we expect
an increase of the electron density in the 2DEG if the LED affects
these doping layers by facilitating further ionisation. In contrast,
the band edge of the screening layer at $z\approx250\,\rm{nm}$ is
situated below the Fermi energy as long as the 2DEG is populated. If
the LED resets the positively charged bottom screening layer, the
electron density of the 2DEG should decrease. Figure~\ref{fig:led}
shows the electron density at $B=0.1\,\rm T$, $T=4.2\,\rm K$,
plotted as a function of gate voltage.
Sweeping the gate bias from $V_{\rm G}=+0.5\,\rm{V}$ to $V_{\rm
G}=-3\,\rm{V}$ (blue curve) and back to $V_{\rm G}=+0.5\,\rm{V}$
(black) creates again a strong density hysteresis. When the
gate-sweep is stopped during accumulation and the IR-LED is switched
on by applying a current of $I=0.1\,\rm mA$, the density decreases
over time, implying that indeed a repopulation of the screening
layer at $z\approx250\,\rm{nm}$ is observed.  Vertical red lines at
voltages $V_{\rm G}=-1.3,-1.06,-0.86\,\rm V$ show that the density
decreases almost towards the depletion value during $15\,\rm
minutes$ of illumination, rather than approaching the value after
waiting (see figure 3). This suggests that some photo-generated
electrons get trapped in the upper screening layer. Indeed, once the
LED is switched off, a slow increase of the electron density is
observed (data not shown), being consistent with the time
dependencies shown in figure 3.

\section{Conclusion}
In conclusion, we investigated gating effects in high-mobility
2DEGs. It is found that a gate-induced ionisation of the doping
layers between top-gate and 2DEG causes a significant hysteresis in
the electron density. Surprisingly, this ionisation has almost no
impact on the 2DEG's mobility at a given electron density. On the
other hand, the depletion of the screening layer beneath the 2DEG is
found to decrease the mobility by more than a factor of 2. This
drastic effect can be explained by a reduced screening of dopants
and a shift of the 2DEG's wave function towards its' lower AlGaAs
boundary. Our finding indicates that for gatable high-mobility
2DEGs, it is important to keep the 2DEG wave function in the center
of the quantum well which could for example be achieved through a
buried back gate. At the same time any possible screening from
X-electrons needs to be (and can be) sacrificed in order to achieve
large density tunability.

\end{document}